# WiMAX or Wi-Fi: The Best Suited Candidate Technology for Building Wireless Access Infrastructure

A.F.M Sultanul Kabir[1], Md.Razib Hayat Khan[2], Abul Ahsan Md.Mahmudul Haque[1] and Mohammad Saiful Islam Mamun[2]
[1]School of Information and Communication Technology, [2]School of Computer and System Sciences
Royal Institute of Technology (KTH) Stockholm Sweden
Email: { afmk, mrhkhan, aamhaque, msimamun}@kth.se

*Abstract:*
**This paper presents a description of the existing wireless technology Wi-Fi and WiMAX, and try to compare Wi-Fi (IEEE 802.11) and WiMAX (IEEE 802.16), with respect to which technology provides a better solution to build a wireless access infrastructure. Each technology is evaluated based on some key characteristics. This paper concludes with a statement of, which technology will be the best and most cost effective solution to end user.**

## I. INTRODUCTION

With the help of many expert communication engineers IEEE has developed various wireless standards in a hierarchical fashion. Some of the deployed wireless standards are: 802.15 (Bluetooth), 802.11 (Wi-Fi), and 802.16 (WiMAX) promoted by WiMAX forum. Recently a new standard, 802.20 for WANs has been proposed, which is currently under development. Each of these IEEE standards has been deployed to fulfill certain criteria and they complement each other.

IEEE 802.11 also known as Wi-Fi standards has had a lot of commercial success, for this reason now the focus of wireless networking shifting to the wide area market. Wi-Fi has been optimized to address the requirements for home or office connectivity but the wide area market is still open to grabs. So to grab the market the low cost wireless which appears is WiMAX, short for Worldwide Interoperability for Microwave Access, is positioned as solution for outdoor and long-range last-mile solutions. Many service providers had adopted this technology as a quick and cheap option to provide connectivity between access points or base stations and their backbone network. The main goal of WiMAX is to provide cheap and fast connectivity of both voice and data communication to remote and difficult terrain locations.

With the increasing market demand for WiMAX, it is now regularly compared with Wi-Fi. While both technologies have some identical technical characteristics, however they are approaching the wireless space from completely different perspectives. The purpose of this paper is to provide a technical and market comparison of Wi-Fi and WiMAX technologies in order to highlight that which technology will be better to build a wireless access infrastructure.

The first part of the paper examines the both of these wireless technology in order to understand both technologies and their underlying concepts. Then, we have discussed some key characteristics to compare the both of these technologies. The last part concludes and presents a conclusion of which will be the best technology to build a wireless access infrastructure.

## II. OVERVIEW OF THE CANDIDATE TECHNOLOGIES.

*2.1 Wi-Fi*
The dream to network PCs and other devices without the cost and complexity of cable infrastructures has driven the rapid growth in the wireless market over the last few years. Wi-Fi is one of the wireless technology which appeared early in the wireless market. Wi-Fi is based on the IEEE 802.11 wireless local area network (WLAN) specification. Actually it was designed to be used indoors at close range for example home user and office environment. The main goal of Wi-Fi technology is to provide service for mobile computing device like laptop. But recently it is used for more services for example consumer device such as televisions, digital cameras, and DVD players.

A user with a mobile computing device such as a laptop, cell phone, or PDA which is Wi-Fi enabled can connect to the global Internet when it is within in range of an access point. The region which is covered by one or more access points is called a hotspot. Hotspots can range from a single room to thousand of square feet's of overlapping hotspots. Wi-Fi can also be used to create a mesh network. Wi-Fi also allows connectivity in peer-to-peer (wireless ad-hoc network) mode, which enables devices to connect directly with each other [1]. This connectivity mode is useful in consumer electronics and gaming applications [1].

Wi-Fi products can use different radio frequencies [2]:
• The 802.11a standard uses 5 GHz in an AP-to-AP interlink.

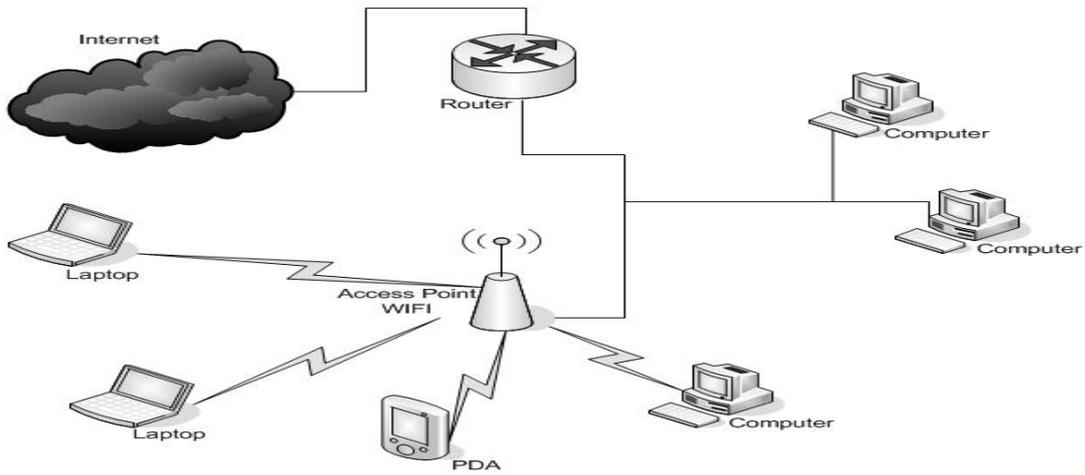

Figure 1: Wi-Fi Network

- The 802.11b and 802.11g standards use 2.4 GHz.

Different frequency bands are used by the 802.11a, 802.11b and 802.11g standards; Different devices using these different frequency bands do not interfere with one another. However, portable devices using different bands cannot communicate with each other, for example an 802.11a radio cannot communicate with an 802.11b radio. The most commonly used standard in the Wireless LAN are the 802.11b and 802.11g standards because of their interoperability and the greater range option that they achieve in the 2.4-GHz band. Each standard also use different types of radio-modulation technology, which is as follows [2]:
• The 802.11b standard uses direct-sequence spread spectrum (DSSS) and supports bandwidth speeds up to 11 Mbps.

• The 802.11a and 802.11g standards use orthogonal frequency division multiplexing (OFDM) and support speeds up to 54 Mbps. Because OFDM is more suitable to outdoor environments and interference, that's why it is commonly used for Wireless LAN infrastructure.

*2.2 WiMAX:*
IEEE standard 802.16, also known as WiMAX, is a technology for last-mile wireless broadband as an alternative to cable and DSL and where the cost is high. It's intended to deliver high speed data communication, and it also has the ability to maintain dedicated links and VoIP services at a reliable and high quality speed.

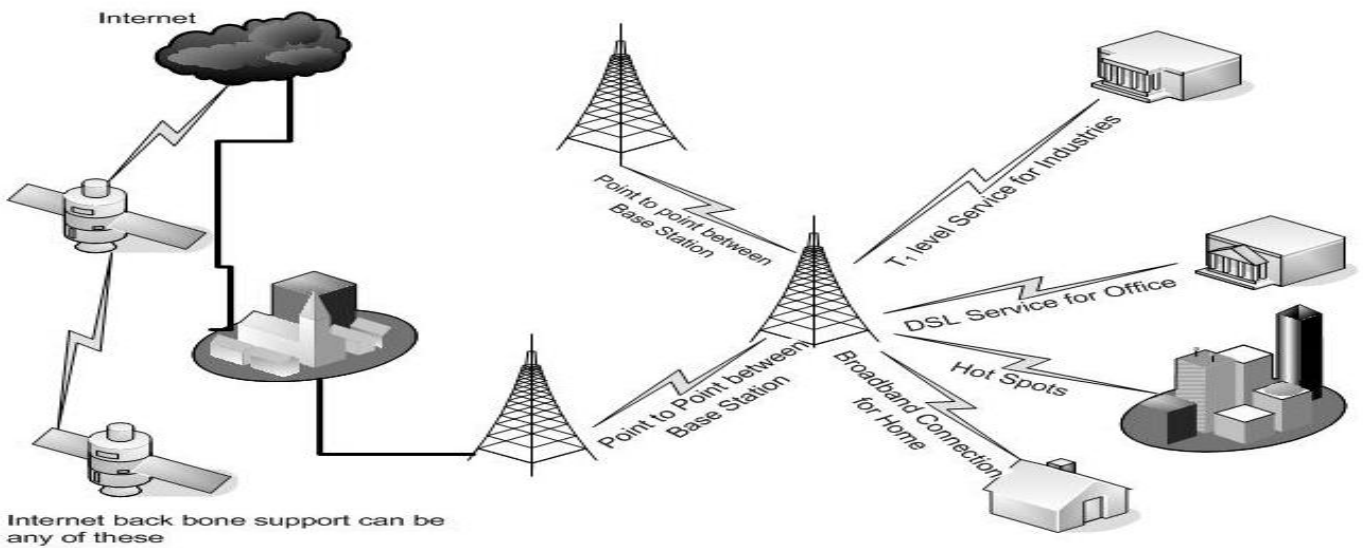

Figure 2: WiMAX Network

Not only it supports "last mile" broadband connectivity to individual home or business locations but also its data rates are comparable with cable and Digital Subscriber Line (DSL) rates. Many telephone companies also desire that WiMAX will be a replacement for their aging legacy wired networks. In fact, it is looked as the wireless replacement for a wired broadband connection.WiMAX has the ability to allow a subscriber to connect to a wireless Internet service provider even when they roam outside their offices or homes.

With the large coverage range and high data transmission rate WiMAX's attributes open the door of the technology to a variety of applications. WiMAX can be used as a backbone for IEEE 802.11 hotspots for connecting to the global world, as well as a subscriber can connect WiMAX enabled mobile devices such as laptops PDA or cell phones directly to WiMAX base stations without using IEEE 802.11.Currently many service providers, providing a DSL or T1/E1 service for a business customer to a relatively remote location or outer suburbs can take several months and the cost associated with it is very high. With the help of WiMAX, a service provider can provide that service in a short time and in a very cost effective way [3].One of the main application of the WiMAX is that it can be used in disaster recovery scenes where the wired networks have broken down. In recent many disasters, WiMAX networks were installed to help in recovery missions [4]. Similarly, WiMAX also be used as a backup links where the traditional wired links breaks. WiMAX mainly operates in two frequency ranges. One is high frequency, which is between 11 – 66 GHz and another one is low frequency, which is sub 11 GHz [3]. Line-of-sight is very essential when operating in the high frequency range. This frequency range allows this wider channel, resulting in very high capacity links. For the low frequency range (sub 11 GHz) non line-of-sight is essential. WiMAX, with a theoretical data rate of 70 Mb/s in 20 MHz channels (2-11GHz spectrum), allows a few hundreds of DSL connections but it operates up to 124Mbps in the 28MHz channel (in 10-66GHz), [5]. The maximum range WiMAX, covered is about 50 km [5]. But in practice this range may be decrease to 20 km and even 8 km when there are obstacles [5].

### III. KEY CHARACTERISTICS OF WIRELESS TECHNOLOGY

This paper focuses on the hypothesis that which wireless technology, WiMAX or Wi-Fi provides a better solution in the wireless access infrastructure. Whether one wireless technology provides a better solution than any other or whether a combination of technologies is needed to create the desired infrastructure. The key characteristics for which the most powerful next generation wireless technology (WiMAX and Wi-Fi) is evaluated in this research paper are: efficiency, maximum range, dependability, security, market issue and mobility. These six key characteristics are the standard issue which will be used to compare these two wireless technologies.

*3.1 Efficiency*
Efficiency of wireless technology is measured in terms of bandwidth and latency. Efficiency is a major issue to determine what type of applications can be run on a network. A less-bandwidth network only feasibly for small application and normally support simple data application for example transferring text files. A higher bandwidth network normally used for big application such as audio and video and many more powerful applications. Another major issue in case of real-time applications like voice is latency which is very much crucial issue. The maximum range of latency should not be more than 20 ms, anything higher than that be warring for establishing echo free wireless network.

*3.2 Maximum Range*
Maximum range is calculated from the obtained distance between the two base stations, and like cell phone another major issue must consider here that the technology must have the capability to support hand-off between base stations without loosing connection from the global world. Maximum coverage range is a major issue, the reason behind that, it determines how long a contiguous wireless area can be? Also, maximum coverage range of wireless technology's is very much crucial according to cost, since operators can reduce their initial capital expenditures if they can give the coverage of the same area with smaller number of base stations.

*3.3 Dependability*
Dependability is defined as how much a wireless technology is dependable to the end user. Whether end user think that is it reliable to use or not? Dependability measure with few important metrics like average number of packet loss, average number of disconnects of calls, and whether the wireless technology is hampered by environmental issues such as line of sight, weather, etc. Dependability is very crucial because some applications may require a reliable connection. If a connection is not dependable, in that case packets may loss and that affect the network for that reason the speed of the network will decrease. This would have certainly impact on the performance of any applications, hence decreasing the applications that will use on the wireless network.

*3.4 Security*
Today's internet is open for all. And user exchange many personal data in internet. So normally end user wants security. Security is obtained from the level of encryption of the data and the authentication of the device is provided by each technology. For many applications such as exchanging bank information require a secure connection to transmit confidential information. Mainly the end user will not want to expose themselves and they also want that the secret information not being viewed by unauthorized individuals. That's why security is needed in wireless connection.

*3.5 Mobility*
Mobility is one of the major issues in case of building wireless access infrastructure. It is the speed of the mobile access point at which the technology can remain connected to the global world without losing packets or service interruption. Naturally, a wireless infrastructure environment needs to be mobile to provide connection to the end user at any place they visit. The network must sustain connection at vehicular speeds.

*3.6 Market comparison*
The last characteristics to consider when evaluating wireless technology is a market. Actually the popularity of any technology is determined by the market. Mainly markets certify a technology whether it is accepted by end user or not. So based upon the market we can decide which technology is most attractive to the wireless world

IV. Wi-Fi VERSUS WiMAX

*4.1 Radio Technology:*
WiMAX differs from Wi-Fi in the radio technology sector. The IEEE 802.11 WLAN standards describe four radio link interfaces that operate mainly in unlicensed radio band having range from 2.4 G to 5 GHz [9]. The WiMAX 802.16a standard released in January 2003 operates between 2 GHz and 11 GHz [9]. The lower frequency bands support Non-line-of-sight (NLOS) for that reason customer unit need not be aligned with base station.

Wi-Fi mainly operates in unlicensed frequency bands, but WiMAX can operate in both licensed and unlicensed spectrum. Within IEEE 802.16a's 2-11 GHz range, four bands are most attractive [9]:

\* Licensed 2.5-GHz MMDS
\* Licensed 3.5-GHz Band
\* Unlicensed 3.5-GHz Band
\* Unlicensed 5 GHz U-NII Band

*4.1.1 Radio transmission Modulation techniques:*
The IEEE 802.11b radio link uses a technique direct sequence spread spectrum that is called complementary coded keying (CCK) for radio transmission [9]. Bit stream is mainly processed by a special coding and modulated with the technique called Quadrature Phase Shift Keying (QPSK). The 802.11a and 802.11g uses the radio link technology 64-channel orthogonal frequency division multiplexing (OFDM) [9]. Here the bit streams is encoded on the 64 sub carriers using Binary Phase Shift Keying (BPSK), Quadrature Phase Shift Keying (QPSK), or one of two levels of Quadrature Amplitude Modulation (16-, or 64- QAM) [9].

The IEEE 802.16a specifies three techniques for radio link [9]:
\* SC-A: Single Carrier Channel.
\* OFDM: 256-Sub-Carrier Orthogonal Frequency Division Multiplexing.
\* OFDM-**A:** 2,048-Sub-Carrier Orthogonal Frequency Division Multiplexing.

*4.2 Efficiency:*
Maximum channel bandwidth for Wi-Fi is 25 MHz for IEEE 802.11b and 20 MHz for either IEEE 802.11a or g networks [9]. The maximum bit rates it's providing is 54 Mbps. Wi-Fi has latency in the range of 50 ms hence little bit higher latency.

In WiMAX, the channel bandwidths are in the range of 1.25 MHz to 20 MHz [9]. Basically there has been lots of confusion regarding the actual bit rate of a WiMAX channel. But many articles give a range in of 70 M or 100 Mbps, basically exact transmission rate depends on the assigned bandwidth of the channel. WiMAX have latency in between the range of 25 to 40 ms, quite considerable range.

Now have a close look at the Bandwidth efficiency of the both technologies. Basically it is measured by the number of bits per second that can be carried on one cycle of radio bandwidth (i.e. bps/Hertz). Lets have a data rates supported on its 25 MHz channel (1 M **to** 11 Mbps), 802.11b have bandwidth efficiency in between 0.04 to 0.44 bps/Hertz [9]. In 802.11 a or g on its 20 MHz have a transmission rate from 6 M to 54 Mbps yields a bandwidth efficiency in between .24 to 2.7 bps/Hertz [9]. In case of WiMAX, for 70-Mbps transmission rate on a 14-MHz radio channel yields bandwidth efficiency up to 5- bits/Hertz [9]. Hence the bandwidth efficiency decreases when the transmission range increases.

*4.3 Maximum Coverage Range:*
OFDM modulation has a high spectral effectiveness that why WiMAX ranges 8 km (NLOS) to 50 km (LOS) [5]. It handles many users who are widely spread out. Mesh topologies and smart antenna techniques can be used to improve the coverage. The OFDM designed for the BWA and main goal is to provide long range transmission. 802.16 is designed for high power OFDM used to maximize coverage up to tens of kilometers [5].

In contrast, IEEE 802.11 standard have a basic CDMA and OFDM approach with a quite different vision. It required very low power consumption of energy that whys it can support very limited range of coverage. It is mainly designed for indoor use. Optimize range of this technology is around 100 meters [5].

*4.4 Security:*
One of the major issues that differentiate from Wi-Fi to WiMAX is security. It's a major issue because it protects transmissions from eavesdropping. But security has been one of the major lacking in Wi-Fi, encryption is optional here. But better encryption techniques are now available some of the different techniques used are [9]:

• Wired Equivalent Privacy (WEP): An RC4-based 40- or 104- bit encryption technique.

- Wi-Fi Protected Access (WPA): A new standard from the Wi-Fi Alliance that uses the 40- or 104-bit WEP key.
- IEEE 802.11i/WPA2**:** It is a IEEE standard which will be based on a more robust encryption technique called the Advanced Encryption Standard.

WiMAX is designed for public network so security is very much crucial here. So all the data that is transmitted in WiMAX network is virtually encrypted. The main encryption technique that is used here is 168-bit Digital Encryption Standard (3DES), the same encryption also used on most secure tunnel VPNs. There are also plan to include the Advanced Encryption Standard (AES) in WiMAX to maximize the security.

*4.5 Mobility Management*
Mobility management is supported by WiMAX. The latest IEEE 802.16e is made for Mobile WiMAX. This standard supports mobile capability with the support of hand-offs capability, mainly for users when they moved between cells. Its support data rates up to 500 kbps, equivalent to the highest speed cellular offerings (e.g. Verizon Wireless' 1xEV-DO service) [9].

Currently mobility management is not supported by Wi-Fi. But recently IEEE has begun to development of a roaming standard for Wi-Fi. However, WLAN switch vendors like Cisco, Aruba, and Airespace have developed their own proprietary hand-off protocols [9].

*4.6 Market Comparison*
Up to this point we have focused on technical issues here we consider, some market issues of these two products. Some market oriented works have been established for Wi-Fi service. The two examples are Wireless ISPs and Wi-Fi mesh networks.

*4.6.1 Wireless ISPs (WISPs)*
The idea behind Wireless ISP (WISP) is to provide an Internet access service using WLAN technology and a shared Internet connection in a public location designated a *hot spot*. TMobile and Wayport are currently providing this type of service [9].But it have two problems, one is technical and another one is business oriented. From a technical viewpoint, to access the internet you have to be within the hot spot. From a business viewpoint, users have to pay monthly basis for the internet then the users have to be in the hot spot always to access the internet which is not a feasible solution. So markets of wireless ISP are in a threat now.

*4.6.2 Wi-Fi Mesh Network*
Wi-Fi mesh networks are mainly used to support public safety applications and also to provide Internet access to end users. However, mesh technologies are not within the range of the Wi-Fi standards.

*4.6.3 WiMAX Market*
The market goals of WiMAX not clear at the moment. But in a sense we can say that the major goal will be broadband wireless access or Wireless DSL. But it will succeed only if it provides lower cost service and also provide some extra features which the other broadband like DSL do not provide. WiMAX compatible chipsets first appeared in late-2004 by the Intel and consumer devices costing $100 or less [9]. But in case of WiMAX, before investing in this field, they have to think and analyze that whether they have enough demand in the market or not.

*4.7 Quality of Service (QoS)*
Wi-Fi is based on a contention based MAC (CSMA/CA).Hence no guaranteed QoS is provided mainly it can support best offer services. The Standard does not permit for different service level for each user. There is a plan to incorporate QoS in the 802.11-e standard. In this standard two operating modes will be included to improve service for voice one is Wi-Fi Multimedia Extensions (WME) and another one is Wi-Fi Scheduled Multimedia (WSM)

QoS in IEEE 802.16 is based on a request/grant protocol. Its support multiple QoS which is build in MAC. It is designed to supports different service levels such as ,T1/E1 for business and best effort to consumer. This protocol support delay sensitive services such as voice and video. The dynamic TDMA based technique allows the suitable support for multicast and broadcast.
In the below the key difference between Wi-Fi and WiMAX is described.

Table:1 Comparison between IEEE 802.11 and IEEE 802.16

|  | 802.11 (Wi-Fi) | 802.16 (WiMAX) |
|---|---|---|
| Primary Application | Wireless LAN | Wireless MAN mainly designed for broadband wireless |
| Range and Coverage | Mainly designed for indoor Optimized for 100 meters No mesh topology is supported | Designed for outdoor NLOS performance Optimized for 50 km Mesh topology is supported |
| Scalability | MAC designed to support tens of user | MAC designed to support thousands of users |
| Frequency Band | Unlicensed Band 2.4 GHz to 5 GHz | Licensed and Unlicensed Band 2 GHz to 11 GHz |
| Channel Bandwidth | On the range from 20-25 MHz | Adjustable range from 1.25 to 20 MHz |
| Bandwidth Efficiency | 0.44 to 2.7 bps/Hz | <=5 bps/Hz |
| Radio Technique | OFDM 64 channels and Direct Sequence Spread Spectrum | OFDM 256 Channels |
| Security | Security is optional here. Better encryption technique like WPA and WEP available now | 3 DES ( 128 bit ) |
| Mobility | In Development phase now | Mobile WiMAX build in to 802.16e |
| QoS | Contention Based MAC (CSMA/CA) QoS is proposed in IEEE 802.11e | Grant Request MAC Mainly designed to support voice and video |

## V. SUMMARY

The key characteristics used to compare the two technologies Wi-Fi and WiMAX were described in section 3: efficiency, maximum coverage range, dependability, measurement of security, mobility issue and market comparison. Using empirical data, we can evaluate which technology has the best capability to design wireless access infrastructure.

• Mainly efficiency is measured from bandwidth and latency. WiMAX (IEEE 802.16e) has the better performance with a maximum bandwidth of 100Mbps and a low latency of 25-40 ms.

• Maximum Coverage area is the measured distance between base stations. And maximum ranges a technology support. We see from the empirical data that WiMAX has better coverage range 50 km than Wi-Fi which supports only 100 meter.

• Estimates of dependability or reliability are best obtained by simulating a network. Both technologies are trying to minimize loss packets and decrease the number of disconnects. Without comparable simulations it is tough to say which technologies support the best dependable solution.

• Security mainly deals with the level of encryption and the device authentication supported by each technology. Each technology has some level of security. Since security feature is very much poor in Wi-Fi. But recently some security feature is added to provide the security more. But WiMAX have a better encryption technique 3 DES. So end user can happily exchange their data in WiMAX.

• Mobility is the speed at which the technology can connect and remain connected. Wi-Fi doesn't support mobility. But 802.16e support mobility. So again we get a better result here for WiMAX.

From this summary it is apparent that WiMAX is a better technology to build wireless access infrastructure.

## VI. CONCLUSION

This paper has studied two emerging wireless standard technologies; Wi-Fi (IEEE 802.11), WiMAX (IEEE 802.16), in terms of how they could be applied to the creation of a wireless access infrastructure. Technical data was collected for both of these technologies and both of these technologies were compared using some key characteristics. It was determined that WiMAX signals the arrival of the next wave of wireless access infrastructure. Limited range and data capability of Wi-Fi helps WiMAX to make a promise of taking high speed wireless out of the coffee shop to the road and everywhere. The main advantage of the WiMAX technology is that it is flexible. Well suited for fixed and nomadic users and also give another advantages by operating in licensed or unlicensed bands, on this paper, it is seen that WiMAX looks like a strong contender to build wireless access infrastructure. Now we will have to wait and see that this technology can capture the market or not. To conclude, it's obvious that the WiMAX standard aim not to replace Wi-Fi in its applications but rather to enhancement it in order to form a modern wireless access infrastructure.